\documentclass{llncs}
\usepackage[utf8]{inputenc}
\usepackage[T1]{fontenc}
\usepackage{booktabs}
\usepackage{cite}
\usepackage{mathtools,amssymb,amsmath}
\usepackage{enumerate}
\usepackage{varioref}
\usepackage[pdfpagelabels=true]{hyperref}
\usepackage{cleveref}
\usepackage{xcolor}
\usepackage{algpseudocode}
\usepackage{algorithm}
\usepackage{listings}
\usepackage{csquotes}
\usepackage{url}
\usepackage{caption}
\usepackage{subcaption}
\newif\iflongversion

\hypersetup{
	linktoc = page,
	pdfpagemode = UseNone,
	colorlinks,
	linkcolor={red!50!black},
	citecolor={blue!50!black},
	urlcolor={blue!80!black}
}

\usepackage{paralist}
\usepackage{ragged2e}
\usepackage{tikz}
\usepackage{pgfplots}
\usepackage{tikz-qtree}
\usepackage[linguistics]{forest}
\usetikzlibrary{arrows}
\usetikzlibrary{backgrounds,decorations.pathreplacing, bending}
\usepackage{breqn}
\definecolor{identifiercolor}{rgb}{.4,.6,.56}
\definecolor{stringcolor}{gray}{0.5}
\definecolor{inactivecolor}{rgb}{0.15,0.15,0.5}

\usepackage{listings}
\begin{document}

    \title{On Catalan Constant Continued Fractions}

    \author{David Naccache\inst{1} \and Ofer Yifrach-Stav\inst{1}} 

    \institute{
    DI\'ENS, \'ENS, CNRS, PSL University, Paris, France\\
45 rue d'Ulm, 75230, Paris \textsc{cedex} 05, France\\
\email{\url{ofer.friedman@ens.fr}}, 
\email{\url{david.naccache@ens.fr}}
}

\maketitle         

\begin{abstract}

The Ramanujan Machine project detects new expressions related to constants of interest, such as $\zeta$ function values, $\gamma$ and algebraic numbers (to name a few).

In particular the project lists a number of conjectures concerning the Catalan constant $G= 0.91596559\ldots$

We show how to generate infinitely many.

We used an ad hoc software toolchain and rather tedious mathematical developments.

Because we do not provide a proper peer-reviewed proof of the relations given here we do not claim them to be theorems.
\end{abstract}
 
\section{Introduction}

The Ramanujan Machine project \cite{rama,ref1,ref2} detects new expressions related to constants of interests, such as $\zeta$ function values, $\gamma$ and various algebraic numbers (to name a few). 

In particular the project lists several of conjectures\footnote{\url{http://www.ramanujanmachine.com/wp-content/uploads/2020/06/catalan.pdf}} concerning the Catalan constant $G= 0.91596559\ldots$

We show how to generate infinitely many.

We used an ad hoc software toolchain and rather tedious mathematical developments.

Because we do not provide a proper peer-reviewed proof of the relations given here we do not claim them to be theorems.

\section{The Initial Conjectures}

Let $a_n=3n^2+(3+4\kappa)n+2\kappa+1$ and $b_n=-2n^2(n+2\kappa)(n+c)$ and consider the continued fraction:

\begin{equation*}\label{eq:gcf}
    Q_{c,\kappa}=a_{0}+{\cfrac {b_{1}}{a_{1}+{\cfrac {b_{2}}{a_{2}+{\cfrac {b_{3}}{a_{3}+{\cfrac {b_{4}}{a_{4}+\ddots \,}}}}}}}}
\end{equation*}

The Ramanujan Project conjectures that:

\begin{center}
  \begin{tabular}{|c|c|c|c|}\hline
~~$c$~~&~~~~$Q_{c,0}$~~~~&~~~~$Q_{c,1}$~~~~ &~~~~$Q_{c,2}$~~~~\\\hline\hline
~~0~~&~~$\frac{1}{2G}$~~        &~~\textcolor{red}{$\frac{2}{2G-1}$}~~&\textcolor{orange}{$\frac{24}{18G-11}$}\\\hline
~~1~~&~~\textcolor{red}{$\frac{2}{2G-1}$}~~      &~~$\frac{4}{2G+1}$~~&\textcolor{blue}{$\frac{16}{6G-1}$}\\\hline
~~2~~&~~\textcolor{orange}{$\frac{24}{18G-11}$}~~   &~~\textcolor{blue}{$\frac{16}{6G-1}$}~~&$\frac{64}{18G+13}$\\\hline
~~3~~&~~$\frac{720}{450G-299}$~~&~~$\frac{288}{90G-31}$~~& ?\\\hline
\end{tabular}
\end{center}  

\section{Notations}

We denote by $n!!$ the semifactorial of, i.e. the product of all the integers from $1$ up to $n$ having the same parity as $n$:

$$n!! = \prod_{k=0}^{\left\lceil\frac{n}{2}\right\rceil - 1} (n-2k) = n (n-2) (n-4) \cdots$$

Because in all the following we will only apply semifactorials to odd numbers, this can simplified as:

$$n!! = \prod_{k=1}^{\frac{n+1}{2}} (2k-1) = n(n-2)(n-4)\cdots 3\cdot 1$$

\section{The Generalized Formulae}

For $\kappa\in\{0,\ldots,6\}$ we have:

$$
\lim_{c \rightarrow \infty} Q_{c,\kappa}-Q_{c-1,\kappa}=2
$$

In the following sections we crunch very smooth constants into factorials and semifactorials. This is done for the sole purpose of saving space and does not imply any specific property.

\subsection{$\kappa=0$}

The generalized form for $\kappa=0$ is:

\begin{equation*}
    Q_{c,0}=\frac{(2c)!}{2(2c-1)!!^2 G-\Delta_{c-1,0}}
\end{equation*}

where:

\[
\Delta_{c,0} =
 \begin{cases}
  1+10c & \mbox{~if~} c < 2 \\
  -2 c (2 c - 1)^3
      \Delta_{c-2,0} + (1 + 2 c + 8 c^2) \Delta_{c-1,0} &  \mbox{~if~} c\geq 2
 \end{cases}
\]

or under an equivalent more compact form:

\[
\Delta_{c,0} =
 \begin{cases}
  1 & \mbox{~if~} c=0 \\
  (2c)!+(2c+1)^2 \Delta_{c-1,0} &  \mbox{~if~} c>0
 \end{cases}
\]

This is easy to check up to any practical rank using the Mathematica code:

\begin{lstlisting}[extendedchars=true,language=Mathematica]
f[x_, {m_, d_}] := m/(d + x);
F1[x_] := (2 x)!;
F2[x_] := 2 (2 x - 1)!!^2;
(*F3[x_]:=If[x==0,1,(2x)!+(2x+1)^2 *F3[x-1]]; or the unfolded form:*)
F3[x_] := 
  If[x < 2, 
   1 + 10 x, (1 + 2 x + 8 x^2) F3[x - 1] - 2 x (2 x - 1)^3 F3[x - 2]];
den = Table[3 n^2 + 3 n + 1, {n, 1, 12000}];
For[c = 1, c < 20, 
 num = Table[-2 n^3 (n + c), {n, 1, 12000}];
 r = 1 + N[(Fold[f, Last@num/Last@den, 
      Reverse@Most@Transpose@{num, den}]), 200];
 e=F1[c]/(Catalan F2[c] - F3[c - 1]);
 Print["Comparison: ", N[{r, e}, 200]];
 c++]
\end{lstlisting}

\remark{At times $\gcd(\gcd(\alpha,\beta),\gcd(\alpha,\gamma))\neq 1$ and $Q$ can be simplified\footnote{A typical example is  $\kappa=0$ and $c=5$ where our formulae predict $3628800/(1285371-1786050G)$, which, after division by 9, yields the reduced value $403200/(142819-198450 G)$}. Because $(2c)!$ and $2((2c-1)!!)^2$ are both smooth, this happens quite frequently.}

\subsection{$\kappa=1$}

The generalized form for $\kappa=1$ is:

\begin{equation*}
    Q_{c,1}=\frac{2 (2 c)!}{2 (2 c - 1)!!^2 G+(2 c - 1) \Delta_{c-1,1}}
\end{equation*}

where:

\[
\Delta_{c,1} =
 \begin{cases}
  1-2c & \mbox{~if~} c < 2 \\
  -2 c (2 c-1) (2 c-3)^2 
      \Delta_{c-2,1} + (-1 - 6 c + 8 c^2) \Delta_{c-1,1} &  \mbox{~if~} c \geq 2
 \end{cases}
\]

As is checked by the following code:

\begin{lstlisting}[extendedchars=true,language=Mathematica]
f[x_, {m_, d_}] := m/(d + x);
F1[x_] := 2 (2 x)!;
F2[x_] := 2 (2 x - 1)!!^2;
F3[x_] := 
  If[x < 2, 
   1 - 2 x, -2 (3 - 2 x)^2 x (-1 + 2 x) F3[
      x - 2] + (-1 - 6 x + 8 x^2) F3[x - 1]];
den = Table[3 n^2 + 7 n + 3, {n, 1, 12000}];
For[c = 1, c < 20,
 num = Table[-2 n^2 (n + 2) (n + c), {n, 1, 12000}];
 r = 3 + N[(Fold[f, Last@num/Last@den, 
      Reverse@Most@Transpose@{num, den}]), 200];
 e=F1[c]/(Catalan F2[c] + (2 c - 1) F3[c - 1]);
 Print["Comparison: ", 
  N[{r, e}, 200]];
 c++]
\end{lstlisting}

\subsection{$\kappa=2$}

The generalized form for $\kappa=2$ is:

\begin{equation*}
Q_{c,2}=\frac{8 (2 c)!}{6 (2c - 1)!!^2 G+(2 c - 1) (2 c - 3)\Delta_{c-1,2}}
\end{equation*}

where:

\[
\Delta_{c,2} =
 \begin{cases}
  1+12c & \mbox{~if~} c < 2 \\
  - 2 c(2 c-5)^2 (-1 + 2 c)
      \Delta_{c-2,2} + (-3 - 14 c + 8 c^2) \Delta_{c-1,2} &  \mbox{~if~} c \geq 2
 \end{cases}
\]

\begin{lstlisting}[extendedchars=true,language=Mathematica]
F1[x_] := 8 (2 x)!;
F2[x_] := 6 (2 x - 1)!!^2;
F3[x_] := 
  If[x < 2, 1+12x, (-3 - 14 x + 8 x^2) F3[x - 1] - 
    2 (5 - 2 x)^2 x (-1 + 2 x) F3[x - 2]];
den = Table[3 n^2 + 11 n + 5, {n, 1, 12000}];
f[x_, {m_, d_}] := m/(d + x);
For[c = 1, c <= 30, num = Table[-2 n^2 (n + 4) (n + c), {n, 1, 12000}];
 r = 5 + N[(Fold[f, Last@num/Last@den, 
      Reverse@Most@Transpose@{num, den}]), 200];
 e = F1[c]/(F2[c] Catalan + (2 c - 1) (2 c - 3) F3[c - 1]);
 Print["Comparison: ", N[{r, e}, 200]];
 c++]
\end{lstlisting}

\subsection{$\kappa=3$}

The generalized form for $\kappa=3$ is:

\begin{equation*}
Q_{c,3}=\frac{(3^3 + 4^3 + 5^3 + 6^3) (2 c)!}{270 (2c - 1)!!^2 G+(2 c - 1) (2 c - 3) (2 c - 5)\Delta_{c-1,3}}
\end{equation*}

where:

\[
\Delta_{c,3} =
 \begin{cases}
  22 c - 31  & \mbox{~if~} c \leq 2 \\
  -2 c(2 c - 1) (2 c - 7)^2 
      \Delta_{c-2,3} + (- 5- 22 c+8 c^2) \Delta_{c-1,3} &  \mbox{~if~} c>2
 \end{cases}
\]

\begin{lstlisting}[extendedchars=true,language=Mathematica]
F1[x_] := 432 (2 x)!; 
F2[x_] := 270 (2 x - 1)!!^2;
F3[x_] := 
  If[x < 2, 
   22 x - 31 , -2 x (2 x - 7)^2 (2 x - 1) F3[
      x - 2] + (8 x^2 - 22 x - 5) F3[x - 1]];
f[x_, {m_, d_}] := m/(d + x);
For[c = 1, c <= 30,
  den = Table[3 n^2 + 15 n + 7, {n, 1, 10000}];
  num = Table[-2 n^2 (n + 6) (n + c), {n, 1, 10000}];
  r = 7 + (Fold[f, Last@num/Last@den, 
      Reverse@Most@Transpose@{num, den}]);
  e = F1[c]/(F2[c] Catalan + (2 c - 1) (2 c - 3) (2 c - 5) F3[c - 1]);
  Print["Comparison: ",N[{r, e}, 200]];
  c++];
\end{lstlisting}

\subsection{$\kappa=4$}

The generalized form for $\kappa=4$ is:

\begin{equation*}
Q_{c,4}=\frac{ 5! 6! (2 c)!}{14\times 15^3 (2c - 1)!!^2 G+(2 c - 1) (2 c - 3) (2 c - 5)(2 c - 7)\Delta_{c-1,4}}
\end{equation*}

where:

\[
\Delta_{c,4} =
 \begin{cases}
  1327 - 10448 c  & \mbox{~if~} c < 2 \\
  -2 c (2 c - 1)(2 c - 9)^2 
      \Delta_{c-2,4} + ( - 7- 30 c +8 c^2) \Delta_{c-1,4} &  \mbox{~if~} c \geq 2
 \end{cases}
\]

\begin{lstlisting}[extendedchars=true,language=Mathematica]
f[x_, {m_, d_}] := m/(d + x);
F1[x_] := 5! 6! (2 x)!;
F2[x_] := 14*15^3 (2 x - 1)!!^2;
F3[x_] := 
  If[x < 2, 
   1327 - 10448 x, -2 x (-9 + 2 x)^2 (-1 + 2 x) F3[
      x - 2] + (-7 - 30 x + 8 x^2) F3[x - 1]];
For[c = 1, c <= 20,
  den = Table[3 n^2 + 19 n + 9, {n, 1, 10000}];
  num = Table[-2 n^2 (n + 8) (n + c), {n, 1, 10000}];
  r = 9 + (Fold[f, Last@num/Last@den, 
      Reverse@Most@Transpose@{num, den}]);
  e = F1[c]/(F2[
        c] Catalan + (2 c - 1) (2 c - 3) (2 c - 5) (2 c - 7) F3[
        c - 1]);
  Print["Comparison: ",N[{r, e}, 200]];
  c++];
\end{lstlisting}

\subsection{$\kappa=5$}

The generalized form for $\kappa=5$ is:

\begin{equation*}
Q_{c,5}=\frac{2\times140^2\times 5!\times (2 c)!}{2 \times 105^3 (2c - 1)!!^2 G+(2 c - 1) \times\cdots\times(2 c - 9)\Delta_{c-1,5}}
\end{equation*}

where:

\[
\Delta_{c,5} =
 \begin{cases}
  -10891 + 
    150002 c  & \mbox{~if~} c < 2 \\
  -2 c (2 c - 1)(2 c - 11)^2 
      \Delta_{c-2,5} + (- 9- 38 c+8 c^2) \Delta_{c-1,5} &  \mbox{~if~} c \geq 2
 \end{cases}
\]

\begin{lstlisting}[extendedchars=true,language=Mathematica]
f[x_, {m_, d_}] := m/(d + x);
F1[x_] := 2*140^2* 5! (2 x)!;
F2[x_] := 2*105^3 (2 x - 1)!!^2;
F3[x_] := 
  If[x < 2, -10891 + 
    150002 x, -2 x (-11 + 2 x)^2 (-1 + 2 x) F3[
      x - 2] + (-9 - 38 x + 8 x^2) F3[x - 1]];
For[c = 1, c <= 20,
  den = Table[3 n^2 + 23 n + 11, {n, 1, 10000}];
  num = Table[-2 n^2 (n + 10) (n + c), {n, 1, 10000}];
  r = 11 + (Fold[f, Last@num/Last@den, 
      Reverse@Most@Transpose@{num, den}]);
  e = F1[c]/(F2[
        c] Catalan + (2 c - 1) (2 c - 3) (2 c - 5) (2 c - 7) (2 c - 
         9) F3[c - 1]);
  Print["Comparison: ",N[{r, e}, 200]];
  c++];
\end{lstlisting}

\subsection{$\kappa=6$}

The generalized form for $\kappa=6$ is:

\begin{equation*}
Q_{c,6}=\frac{12\times 7!!\times 10!\times (2 c)!}{2\times 7!!\times 9!!\times 11!! (2c - 1)!!^2 G+(2 c - 1) \times\cdots\times(2 c - 11)\Delta_{c-1,6}}
\end{equation*}

where:

\[
\Delta_{c,6} =
 \begin{cases}
  1167809 - 
    23021852 c  & \mbox{~if~} c < 2 \\
  -2 c (2 c - 1)(2 c - 13)^2 
      \Delta_{c-2,6} + (- 11- 46 c +8 c^2) \Delta_{c-1,6} &  \mbox{~if~} c \geq 2
 \end{cases}
\]

\begin{lstlisting}[extendedchars=true,language=Mathematica]
f[x_, {m_, d_}] := m/(d + x);
F1[x_] := 12*7!!*10! (2 x)!;
F2[x_] := 2*11!!*9!!*7!! (2 x - 1)!!^2;
F3[x_] := 
  If[x < 2, 
   1167809 - 
    23021852 x, -2 x (-13 + 2 x)^2 (-1 + 2 x) F3[
      x - 2] + (-11 - 46 x + 8 x^2) F3[x - 1]];
For[c = 1, c <= 20,
  den = Table[3 n^2 + 27 n + 13, {n, 1, 10000}];
  num = Table[-2 n^2 (n + 12) (n + c), {n, 1, 10000}];
  r = 13 + (Fold[f, Last@num/Last@den, 
      Reverse@Most@Transpose@{num, den}]);
  e = F1[c]/(F2[
        c] Catalan + (2 c - 1) (2 c - 3) (2 c - 5) (2 c - 7) (2 c - 
         9) (2 c - 11) F3[c - 1]);
  Print["Comparison: ",N[{r, e}, 200]];
  c++];
\end{lstlisting}

\subsection{$\kappa=7$ \& Beyond}

Nothing fundamentally different is expected happen as we increase $\kappa$. In other words we should keep getting limits of the form:

\begin{equation*}
Q_{c,\kappa}=\frac{\sigma_{\mbox{\scriptsize num},\kappa}\times (2 c)!}{\sigma_{\mbox{\scriptsize den},\kappa} \times(2c - 1)!!^2 G+(2 c - 1) \times\cdots\times(2(c-\kappa)+1))\Delta_{c-1,\kappa}}
\end{equation*}

Where $\sigma_{\mbox{\scriptsize num},\kappa},\sigma_{\mbox{\scriptsize den},\kappa}$ are composed of a (very) smooth sub-factor times a few potentially larger primes. The polynomials that intervene in the definition of  $\Delta_{c,\kappa}$ are :

$$8c^2+(2-8\kappa)c-2\kappa+1\mbox{~~and~~}-2 c (2 c-1) (2(c-\kappa)-1)^2 $$

For instance we have: $\sigma_{\mbox{\scriptsize num},7}=1024$, $\sigma_{\mbox{\scriptsize den},7}=429$ and in $\Delta_{c,7}$ the $c < 2$ case is defined by:

$$\frac{2258335679 c}{35 \times11!!^2} -\frac{176673487}{70\times 11!!^2}$$

We have good reasons to conjecture that the ratios of the coefficients $\sigma_{\mbox{\scriptsize num},\kappa},\sigma_{\mbox{\scriptsize den},\kappa}$ appearing in front of \texttt{F1} and \texttt{F2} in our code, i.e. $2/2, 8/6, 432/270,\ldots$ satisfy\footnote{With $\kappa=0$ being an exception that does not fit this general rule.}: 

$$\frac{\sigma_{\mbox{\scriptsize num},\kappa}}{\sigma_{\mbox{\scriptsize den},\kappa}}=\frac{4^{\kappa - 1}}{(2 \kappa - 1)C_{\kappa - 1}}\mbox{~where~}C_{\ell}\mbox{~is the~}\ell\mbox{-th Catalan number.}$$

That is:

$$\frac{\sigma_{\mbox{\scriptsize num},\kappa}}{\sigma_{\mbox{\scriptsize den},\kappa}}=\left\{\frac{1}{2},1,\frac{4}{3},\frac{8}{5},\frac{64}{35},\frac{128}{63},\frac{512}{231},\frac{1024}{429},\ldots\right\}$$

The conjecture that the $\sigma_{\mbox{\scriptsize den},\kappa}$ and $\sigma_{\mbox{\scriptsize num},\kappa}$ are given by the function $1/T_s^k$ of \cite{tsk} (cf. Table 1 of \cite{tsk}) is backed by the fact that the same type of ratios between $\alpha$ and $\gamma$ in $Q=\alpha/(\beta+\gamma G))$ can also be observed for polynomials that do not belong to the $\kappa$ family. For instance\footnote{modulo a possible sign flip for the very first values}:

\begin{table}[H]
    \centering
    \begin{tabular}{|c|c|c|c|c|c|}\hline
~~$\delta$~~&~~$\epsilon$~~&~~$\eta$~~&~~$\tau$~~& relation \\\hline\hline
        15    &  15        &   2    &    4   & $2^{2c+2} \gamma =\phantom{1}3\alpha(2 c - 5) C_{c - 1}$ \\\hline 
        19    &  21        &   2    &    6   & $2^{2c+ 3}\gamma =\phantom{1}5\alpha(2 c - 7) C_{c - 1}$ \\\hline
        19    &  25        &   4    &    4   & $2^{2c+4} (2c-3) \gamma=\phantom{1}9\alpha(2c-7)  (2c-5) C_{c-1} $ \\\hline
        23    &  35        &   4    &    6   &~~$2^{2c+5} (2c-3) \gamma =15\alpha(2c-7) (2c-9)  C_{c-1}$~~\\\hline
        \end{tabular}
    \caption{Ratios between $\alpha$ and $\gamma$ for various example continued fractions.}
    \label{tab:ratios}
\end{table}

The above provides an algorithmic way to compute generalized relations.

\paragraph{Step 1} For a given $\kappa$, generate\footnote{By any analytic or computational manner.} 2 limits, e.g. for $c=1$ and $c=2$. 

\paragraph{Step 2} We have four unknowns to determine: the coefficients $\sigma_{\mbox{\scriptsize num},\kappa},\sigma_{\mbox{\scriptsize den},\kappa}$ in front of \texttt{F1} and \texttt{F2} and the parameters $A,B$ in the equation $Ac+B$ at the ``if $c<2$'' part of $\Delta$. The ratio of $\sigma_{\mbox{\scriptsize num},\kappa}$ and $\sigma_{\mbox{\scriptsize den},\kappa}$ being known, this reduces the number of unknowns to three.

\paragraph{Step 3} Arbitrarily set $\sigma_{\mbox{\scriptsize num},\kappa}=1$ and push $\sigma_{\mbox{\scriptsize den},\kappa}$ into $A$ and $B$. We are hence left with two rational unknowns which can be found by solving a system of two linear equations given by the two limits we started with.\smallskip

Once this ``bootstraping'' achieved, we can keep generating limits using the generalized formula for that specific $\kappa$ but for any desired rank $c$.

\subsection{Other $G$ Relations of the Ramanujan Machine Project}

The same techniques generate many relations unlisted by the Ramanjuan Project. We give a few examples that we computed (or re-computed) up to rank 6\footnote{There is no intrinsic difficulty to keep computing those constants, we limit ourselves to 7 per category for space reasons.} in the following tables where:
$$Q=\frac{\alpha}{\beta+\gamma G},\mbox{~~~~}a_n=3 n^2 + \delta n + \epsilon,\mbox{~~and~~}
b_n=-2n(n+\tau)(n+\eta)(n+\mu)$$

\begin{table}[h]
{\scriptsize
\setcounter{subtable}{3} 
\begin{subtable}[]{0.45\textwidth}
\centering
\begin{tabular}{|c|c|c|c|}\hline
~~$\mu$~~&~~~~~~~~$\alpha$~~~~~~~~&$~~~~~~~~\beta$~~~~~~~~&~~~~~~~~$\gamma$~~~~~~~~\\\hline
 0&1&0&2\\\hline
 1&2&-1&2\\\hline
 2&24&-11&18\\\hline
 3&720&-299&450\\\hline
 4&40320&-15371&22050\\\hline
 5&403200&-142819&198450\\\hline
 6&53222400&-17684299&24012450\\\hline
\end{tabular}
    \caption{$\delta=3$, $\epsilon=1$, $\tau=0$, $\eta=0$}
    \label{tab:3100}
\end{subtable}}
\hfill
{\scriptsize
\begin{subtable}[]{0.45\textwidth}
    \centering
\begin{tabular}{|c|c|c|c|}\hline
~~$\mu$~~&~~~~~~~~$\alpha$~~~~~~~~&$~~~~~~~~\beta$~~~~~~~~&~~~~~~~~$\gamma$~~~~~~~~\\\hline
0&2&-1&2\\\hline
1&4&1&2\\\hline
2&16&-1&6\\\hline
3&288&-31&90\\\hline
4&11520&-1373&3150\\\hline
5&89600&-10891&22050\\\hline
6&9676800&-1167809&2182950\\\hline
\end{tabular}
\caption{$\delta=7$, $\epsilon=3$, $\tau=0$, $\eta=2$}
\label{tab:7302}
\end{subtable}}
\\
{\scriptsize
\begin{subtable}[]{0.45\textwidth}
\centering
\begin{tabular}{|c|c|c|c|}\hline
~~$\mu$~~&~~~~~~~~$\alpha$~~~~~~~~&$~~~~~~~~\beta$~~~~~~~~&~~~~~~~~$\gamma$~~~~~~~~\\\hline
0&24&-11&18\\\hline
1&16&-1&6\\\hline
2&64&13&18\\\hline
3&384&1&90\\\hline
4&3072&-121&630\\\hline
5&51200&-2839&9450\\\hline
6&4300800&-269803&727650\\\hline
\end{tabular}
\caption{$\delta=11$, $\epsilon=5$, $\tau=0$, $\eta=4$}
\label{tab:11504}
\end{subtable}}
\hfill
{\scriptsize
\begin{subtable}[]{0.45\textwidth}
\centering
\begin{tabular}{|c|c|c|c|}\hline
~~$\mu$~~&~~~~~~~~$\alpha$~~~~~~~~&$~~~~~~~~\beta$~~~~~~~~&~~~~~~~~$\gamma$~~~~~~~~\\\hline
0&4&-5&6\\\hline
1&8&3&-2\\\hline
2&32&5&2\\\hline
3&192&13&18\\\hline
4&4608&133&450\\\hline
5&230400&1909&22050\\\hline
6&2150400&-8419&198450\\\hline
\end{tabular}
\caption{$\delta=11$, $\epsilon=9$, $\tau=2$, $\eta=2$}
\label{tab:11922}
\end{subtable}}
\hfill
{\scriptsize
\begin{subtable}[]{0.45\textwidth}
\centering
\begin{tabular}{|c|c|c|c|}\hline
~~$\mu$~~&~~~~~~~~$\alpha$~~~~~~~~&$~~~~~~~~\beta$~~~~~~~~&~~~~~~~~$\gamma$~~~~~~~~\\\hline
0&720&-299&450\\\hline
1&288&-31&90\\\hline
2&384&1&90\\\hline
3&2304&389&450\\\hline
4&18432&419&3150\\\hline
5&61440&-791&9450\\\hline
6&737280&-20989&103950\\\hline
\end{tabular}
\caption{$\delta=15$, $\epsilon=7$, $\tau=0$, $\eta=6$}
\label{tab:15706}
\end{subtable}}
\hfill
{\scriptsize
\begin{subtable}[]{0.45\textwidth}
\centering
\begin{tabular}{|c|c|c|c|}\hline
~~$\mu$~~&~~~~~~~~$\alpha$~~~~~~~~&$~~~~~~~~\beta$~~~~~~~~&~~~~~~~~$\gamma$~~~~~~~~\\\hline
0&48&-79&90\\\hline
1&32&19&-18\\\hline
2&128&17&-6\\\hline
3&768&77&18\\\hline
4&2048&129&90\\\hline
5&61440&2467&3150\\\hline
6&1228800&31327&66150\\\hline
\end{tabular}
\caption{$\delta=15$, $\epsilon=15$, $\tau=4$, $\eta=2$}
\label{tab:151543}
\end{subtable}}
\\
{\scriptsize
\begin{subtable}[]{0.45\textwidth}
\begin{tabular}{|c|c|c|c|}\hline
~~$\mu$~~&~~~~~~~~$\alpha$~~~~~~~~&$~~~~~~~~\beta$~~~~~~~~&~~~~~~~~$\gamma$~~~~~~~~\\\hline
0&1440&-2813&3150\\\hline
1&576&443&-450\\\hline
2&768&127&-90\\\hline
3&4608&383&-90\\\hline
4&36864&2693&450\\\hline
5&122880&6563&3150\\\hline
6&294912&11497&9450\\\hline
\end{tabular}
\caption{$\delta=19$, $\epsilon=21$, $\tau=2$, $\eta=6$}
\label{tab:192126}
\end{subtable}}
\hfill
{\scriptsize
\begin{subtable}[]{0.45\textwidth}
\begin{tabular}{|c|c|c|c|}\hline
~~$\mu$~~&~~~~~~~~$\alpha$~~~~~~~~&$~~~~~~~~\beta$~~~~~~~~&~~~~~~~~$\gamma$~~~~~~~~\\\hline
0&192&-569&630\\\hline
1&128&253&-270\\\hline
2&512&-25&54\\\hline
3&3072&179&-18\\\hline
4&8192&487&54\\\hline
5&81920&3983&1350\\\hline
6&327680&12583&7350\\\hline
\end{tabular}
\caption{$\delta=19$, $\epsilon=25$, $\tau=4$, $\eta=4$}
\label{tab:192544}
\end{subtable}}
\\
\begin{center}
{\scriptsize
\begin{subtable}[]{0.45\textwidth}
\begin{tabular}{|c|c|c|c|}\hline
~~$\mu$~~&~~~~~~~~$\alpha$~~~~~~~~&$~~~~~~~~\beta$~~~~~~~~&~~~~~~~~$\gamma$~~~~~~~~\\\hline
0&1920&-8599&9450\\\hline
1&768&2909&-3150\\\hline
2&1024&-379&450\\\hline
3&2048&43&30\\\hline
4&49152&1919&-90\\\hline
5&163840&6789&450\\\hline
6&393216&14755&3150\\\hline
\end{tabular}
\caption{$\delta=23$, $\epsilon=35$, $\tau=4$, $\eta=6$}
\label{tab:233546}
\end{subtable}}
\end{center}
\end{table}

\begin{table}[]
\centering
\begin{tabular}{|c|c|c|c|c|c|c|c|}\hline
~~$\delta$~~&~~$\epsilon$~~&~~$\tau$~~&~~$\eta$~~&~~$\mu$~~&~~~~~~~~$\alpha$~~~~~~~~&$~~~~~~~~\beta$~~~~~~~~&~~~~~~~~$\gamma$~~~~~~~~\\\hline
9&7&1&1&1&1&2&-2\\\hline
13&13&1&1&3&6&17&-18\\\hline
15&19&2&2&2&8&-49&54\\\hline
17&19&1&1&5&120&419&-450\\\hline
17&23&1&3&3&12&83&-90\\\hline
19&29&2&2&4&32&-411&450\\\hline
21&33&1&3&5&240&2893&-3150\\\hline
\end{tabular}
\caption{Sporadic case examples}
\label{tab:spor}
\end{table}

\begin{table}[]
    \centering
\begin{tabular}{|c|c|c|c|c|}\hline
~~23,39,2,2,6~~&~~23,43,2,4,4~~&~~25,31,1,1,9~~&~~25,43,1,3,7~~&~~25,47,1,5,5~~\\\hline
~~25,51,3,3,5~~&~~21,25,1,1,7~~&~~21,37,3,3,3~~&~~27,57,2,4,6~~&~~27,61,4,4,4~~\\\hline
~~29,37,1,1,11~~&~~29,53,1,3,9~~&~~31,59,2,2,10~~&~~33,43,1,1,13~~&~~37,49,1,1,15~~\\\hline
\end{tabular}
\caption{Other sporadic convergence examples, entries are given in the following order: $\delta$,$\epsilon$,$\tau$,$\eta$,$\mu$.}
\label{tab:spor2}
\end{table}

\begin{table}[]
    \centering
\begin{tabular}{|c|c|c|}\hline
~~~~~~$\delta$~~~~~~&~~~~~~$\eta$~~~~~~&~~~~~~$\tau$~~~~~~\\\hline\hline
    $4i+7$ & $0$ & $2i+2$\\\hline
    $4i+11$ & $2$& $2i+2$\\\hline
    $4i+19$&$4$&$4i+4$\\\hline
    $4i+27$&$6$&$2i+6$\\\hline
    $4i+35$&$8$&$2i+8$\\\hline
\end{tabular}
\caption{Examples of families. Running $\mu$, $i=0,1,\ldots$, $\epsilon=(\eta+1)(\tau+1)$.}\label{newones}
\end{table}

To keep generating such examples fix an integer bound $\ell$, set $\mu$ to any arbitrary positive integer, run $i=3,\ldots,\ell$, $j=0,\ldots,\lfloor i/2 \rfloor+1$. Then the polynomials $a_n=j (2 i - j + 2) + (4 i + 3) n + 3 n^2$ and $b_n=-2 n (n + j - 1)(n + 2 i - j + 1)(n + \mu)$ will result in $Q$ values of the form $\alpha/(\beta+G\gamma)$. Increments of $i$ by rational steps (e.g. $\frac{1}{2}$ instead of $1$) generate additional relations. Table \ref{last} gives a few examples for $\mu=3$ (chosen arbitrarily as an example). For $0\leq i\leq 2$ we have $Q=0$.\smallskip

This is one example, as there are other similar families: amongst the converging examples over a running variable $\mu$ and $i=0,1,\ldots$, three are given in table \ref{newones} where  $\epsilon=(\eta+1)(\tau+1)$.

Some relations do not belong to families and are sporadic (e.g. table \ref{tab:spor}).

\subsubsection{Note:} The same polynomials also yield relations not involving $G$, those are provable by standard techniques. For instance for $\delta=\eta=15$ and denoting $b_n=-2n(n+4)\phi(n)$ we get for all $i$ the results in Table \ref{tab:nog}.

\begin{table}[]
\centering
\begin{tabular}{|c|c|}\hline
$\phi(n)$&~~converges to $Q$~~             \\\hline\hline
 $(n-1)(n+i)$            &    $15$             \\\hline
 $(n-2)(n+i)$            &   $\frac{10 i + 505}{33}$\\\hline
~~$(n-3)(n+2i+1)$~~&    $\frac{25 (421 + 40 i)}{651 + 16 i}$\\\hline
~~$(n-3)(n+2i)$~~&   $\frac{25 (401 + 40 i)}{643 + 16 i}$\\\hline
~~$(1+n/2)(2n+2j-1)$~~&    
$\begin{cases}
  12 & \mbox{~for~} j=2 \\
  10 & \mbox{~for~} j=3 \\
  0 & \mbox{~for~} j\geq 4  \\
 \end{cases}$\\\hline
\end{tabular}
\caption{Examples not involving $G$}
\label{tab:nog}
\end{table}
\subsection{Further extensions}

Unless we missed something fundamental in the other relations involving $G$ of the Ramanujan Machine project, the same type of generalization should \textsl{in principle} apply to the other $G$ relations (given online as of October 2022).

\bibliographystyle{abbrv}
\bibliography{biblio.bib}

\begin{thebibliography}{1}

\bibitem{tsk}
I.~Cação, M.~Falcão, and H.~Malonek.
\newblock Hypercomplex polynomials, vietoris’ rational numbers and a related
  integer numbers sequence.
\newblock {\em Complex Analysis and Operator Theory}, 11, 06 2017.

\bibitem{ref1}
N.~B. David, G.~Nimri, U.~Mendlovic, Y.~Manor, and I.~Kaminer.
\newblock {On the Connection Between Irrationality Measures and Polynomial
  Continued Fractions}, 2021.

\bibitem{rama}
G.~Raayoni, S.~Gottlieb, Y.~Manor, G.~Pisha, Y.~Harris, U.~Mendlovic, D.~Haviv,
  Y.~Hadad, and I.~Kaminer.
\newblock {Generating Conjectures on Fundamental Constants with the Ramanujan
  Machine}.
\newblock {\em Nature}, 590(7844):67--73, Feb 2021.

\bibitem{ref2}
G.~Raayoni, G.~Pisha, Y.~Manor, U.~Mendlovic, D.~Haviv, Y.~Hadad, and
  I.~Kaminer.
\newblock {The Ramanujan Machine: Automatically Generated Conjectures on
  Fundamental Constants}.
\newblock {\em CoRR}, abs/1907.00205, 2019.

\end{thebibliography}
\appendix
\begin{table}[]
\centering
{\scriptsize
\begin{tabular}{|c|c|c|c|c|}\hline
~~$i$~~&~~$j$~~&$\alpha$&$\beta$&$\gamma$\\\hline\hline
0&1&-720&-299&450\\\hline\hline
1&1&288&31&-90\\\hline\hline
2&1&-384&1&90\\\hline
2&2&-6&1&0\\\hline\hline
3&1&-2304&389&450\\\hline
3&2&-6&1&0\\\hline\hline
4&1&18432&-419&-3150\\\hline
4&2&210&-19&0\\\hline
4&3&-4608&383&-90\\\hline\hline
5&1&61440&791&-9450\\\hline
5&2&630&-41&0\\\hline
5&3&-12288&1145&-630\\\hline\hline
6&1&737280&20989&-103950\\\hline
6&2&1386&-71&0\\\hline
6&3&122880&-13079&9450\\\hline
6&4&378&-11&0\\\hline\hline
7&1&-72253440&-2647279&9459450\\\hline
7&2&-2574&109&0\\\hline
7&3&7372800&-884203&727650\\\hline
7&4&297&-7&0\\\hline\hline
8&1&231211008&9547469&-28378350\\\hline
8&2&-858&31&0\\\hline
8&3&80281600&-10675439&9459450\\\hline
8&4&858&-17&0\\\hline
8&5&-9830400&-1833409&2182950\\\hline\hline
9&1&-45779779584&-2016587711&5306751450\\\hline
9&2&-6630&209&0\\\hline
9&3&2312110080&-336233167&312161850\\\hline
9&4&-117&2&0\\\hline
9&5&-963379200&-272007887&312161850\\\hline\hline
10&1&11902742691840&543876944201&-1310767608150\\\hline
10&2&9690&-271&0\\\hline
10&3&457797795840&-71995419827&68987768850\\\hline
10&4&1530&-23&0\\\hline
10&5&-9248440320&-3600327811&4058104050\\\hline
10&6&90&-1&0\\\hline\hline
11&1&12469539962880&581663428937&-1310767608150\\\hline
11&2&-13566&341&0\\\hline
11&3&-7935161794560&1337393123657&-1310767608150\\\hline
11&4&969&-13&0\\\hline
11&5&-122079412224&-61822135475&68987768850\\\hline
11&6&-102&1&0\\\hline\hline
12&1&-5087572304855040&-239899940677247&512510134786650\\\hline
12&2&18354&-419&0\\\hline
12&3&124695399628800&-22357818254809&22283049338550\\\hline
12&4&2394&-29&0\\\hline
12&5&31740647178240&20090629170649&-22283049338550\\\hline
12&6&-114&1&0\\\hline
12&7&-732476473344&5376960927599&-5863960352250\\\hline\hline
13&1&~~502652143719677952~~&~~23808008825309473~~&~~-48688462804731750~~\\\hline
13&2&-4830&101&0\\\hline
13&3&50875723048550400&-9645671177722733&9737692560946350\\\hline
13&4&-1449&16&0\\\hline
13&5&-498781598515200&-383233413631771&423377937432450\\\hline
13&6&126&-1&0\\\hline
13&7&-38088776613888&388070083677979&-423377937432450\\\hline
\end{tabular}}
\caption{Examples converging to $Q$ values involving $G$. Polynomials are $a_n=j (2 i - j + 2) + (4 i + 3) n + 3 n^2$ and $b_n=-2 n (n + j - 1) (n + 2 i - j + 1) (n + \mu)$. $\mu=3$ was chosen arbitrarily for the sake of the example. $\forall i$, $j=0\Rightarrow (\alpha,\beta,\gamma)=(0,0,1)$ (omitted).}
\label{last}
\end{table}
\end{document}